\documentclass[pre,twocolumn,showpacs,preprintnumbers,amsmath,amssymb]{revtex4}

\usepackage{graphicx}
\usepackage{dcolumn}
\usepackage{bm}

\begin{document}

\newcommand{\beq}{\begin{equation}}
\newcommand{\eeq}{  \end{equation}}
\newcommand{\bea}{\begin{eqnarray}}
\newcommand{\eea}{  \end{eqnarray}}
\newcommand{\bit}{\begin{itemize}}
\newcommand{\eit}{  \end{itemize}}

\title{Loschmidt Echo and the Local Density of States}
\author{Natalia Ares}
\author{Diego A. Wisniacki}
 \email{wisniacki@df.uba.ar}
\affiliation{Departamento de F\'{\i}sica ''J. J. Giambiagi'', FCEN, UBA,
1428 Buenos Aires, Argentina.}
\date{\today}
\begin{abstract}

Loschmidt echo (LE)  is a measure of  reversibility and
sensitivity to perturbations of quantum evolutions.  For weak perturbations 
its decay rate is given by the width of the local density 
of states (LDOS).  When the perturbation is strong enough,  it has been 
shown in chaotic systems that its decay is  dictated by the classical Lyapunov exponent. 
However, several recent studies have shown an unexpected 
non-uniform decay rate as a function of the perturbation strength instead of that 
Lyapunov decay. Here we study the systematic behavior of this regime in perturbed 
cat maps. 
We show that some perturbations produce coherent oscillations in the width of LDOS 
that imprint clear signals of the perturbation in LE decay. We also show that if the perturbation 
acts in a small region of phase space (local perturbation) the effect is magnified and 
the decay is given by the width of the LDOS.

\end{abstract}

\maketitle

\section{Introduction}
Reversibility and sensibility to perturbations of quantum systems are at the heart of fields of research as vast as quantum computation, quantum control and coherent transport \cite{nielsen&chuang:qc,rice&zhao,datta}. The interest in these subjects has greatly increased  due to the development of experimental techniques that enable the manipulation of a great number of quantum systems from photons to mesoscopic devices \cite{haroche,meso}.
\vskip 5pt
The suitable magnitude for measuring the stability of the quantum motion, as well as its irreversibility, is the Loschmidt Echo (LE) \cite{Peres,Jalabert-Pastawski}, defined as,
\begin{equation}
M(t)=\vert\langle \psi _0 \vert exp(iHt)exp(-iH_0t)\vert \psi_0 \rangle\vert^{2}
\end{equation}
where $\psi_0$ is the initial state, $H_0$ the nonperturbed Hamiltonian and $H$ the slightly perturbed one. For convenience, we set $\hbar=1$ throughout the paper. The initial  state $\psi_0$ is usually a Gaussian wave packet and
it frequently refers to an average taken over an ensemble of initial states randomly localized in 
phase space.
The LE has been extensively studied in recent years and different time and perturbation 
strength regimes were shown to exist \cite{jacquod,prosen-rev,jacquod-rev,wisniacki1}. As a function of time, this magnitude has three well known regimes. For very short times, it is parabolic or Gaussian, as the perturbation theory is valid to first order \cite{wisniacki2}. This transient regime is followed by a decay that is typically exponential in chaotic systems \cite{jacquod}. Finally, the LE finds a long-time saturation at values inversely proportional to the effective size of the Hilbert space.

As a function of the strength of the perturbation, the decay of the LE has mainly tree different 
behaviors  \cite{Jalabert-Pastawski, jacquod}. 
When the perturbation is very small, in which a typical matrix 
element $W$ of the perturbation is smaller than the mean level 
spacing $\Delta$, the decay is  Gaussian until $M$  reaches its asymptotic values.  
If $W>\Delta$, this regime 
has an exponential decay, with decay rate given by the width $\sigma$ of the 
local density of states (LDOS). This is usually called Fermi 
golden rule regime (FGR).   Finally, when $\sigma>\lambda$
 , with $\lambda$ the mean 
Lyapunov exponent of the classical system,
the regime becomes independent of the perturbation and the decay rate is given by $\lambda$. 
Although this Lyapunov regime seems to be universal as the intensive numerical studies
have shown in the literature \cite{prosen-rev,jacquod-rev},
recent works have found a non-uniform behavior of the decay rate as a function of the perturbation strength. This was observed in an echo spectroscopy  experiment on ultracold atoms in optical billiards and in theoretical studies of the kicked rotator, 
the sawtooth map and Josephson flux qubits  \cite{coldatoms,Casati 2004,wang, Li 2005, Pozzo 2007}. 
Similar qualitative behavior was shown in  Ref. \cite{Richter 2008} for local perturbations.
In this case the authors found  an oscillating regime of the decay of the LE around the value of the classical escape rate. 

Our main aim in this paper is to study the non-uniform decay rate  of the LE 
in the strong perturbation regime mentioned before.
We show that the origin of such behavior is related to the interplay between the width LDOS
and the Lyapunov exponent. 
The LDOS is a measure of the action of a perturbation over  a system and it gives  
the spread of the states of the unperturbed Hamiltonian in the bases of the perturbed one. 
In general, the width of the LDOS has a quadratic growth when the strength of the perturbation is small and, for increasing 
strength, the LDOS has a constant plateau $\bar{\sigma}$ or an oscillating behavior around
$\bar{\sigma}$. 
This regime, which does not present 
a steady growth, is related to the finite number of states of the Hilbert space linked by the perturbation
\cite{doron0}. 
We show that if $\bar{\sigma}\gg\lambda$ we obtain
the expected result that the decay rate of the LE is given by  $\lambda$ when the 
perturbation is strong enough. 
However,  if  $\bar{\sigma} \leq\lambda$, the non-uniform behavior of the width of the LDOS around  $\bar{\sigma}$ seems to be imprinted in the decay rate of the LE, that is, the decay of the echo shows fluctuations that can mask the Lyapunov decay.  This unexpected result can be understood using the semiclassical
dephasing representation of the LE \cite{Vanicek 2003}. 
We also consider local perturbations, that is, the case where the perturbation is applied to a window of the 
positions. These perturbations allow to change the width of the
LDOS while the Lyapunov exponent remains fixed as  the region of the applied
perturbation is varied. 
Moreover,  the use of local perturbations is motivated by  recent semiclassical work of  Ref. \cite{Richter 2008} and by the fact that local perturbations are more amenable for laboratory experiments. 
 In this case we show that the smaller the slice of the perturbed region is, the more similar the decay rate of the LE and the width of the LDOS are. 
  
The model  that we use in our study is  the cat map, a paradigmatic system in quantum chaos that 
is attracting increasing interest in the area of quantum computation 
\cite{miquel,qc-cat}.  This system is completely chaotic and is very well suited to our work. 
It can be  perturbed in different ways and  the value of $\lambda$ can be chosen at will. Moreover, the perturbations can be easily applied to all 
(global) or  a section (local) of the phase space. 
  
The paper is organized as follows.  Section \ref{sec:system} describes the main characteristics of the classical and quantum cat maps .  We introduce the perturbations used in our work and  the LDOS, a measure of the perturbation action over the system.  In Section  \ref{sec:LE}  we show the results of the decay of the LE for cat maps with different Lyapunov exponents using different perturbations.  In this section we show the relation between the decay of the LE  and the width of the LDOS. In Section  \ref{sec:LELP} we make a similar study  but for local perturbations.  Finally, Section \ref{sec:FR}  is devoted to final remarks and conclusions. 

\section{Perturbed Cat Maps}
\label{sec:system}

Torus maps are canonical examples of classical chaotic systems. In this work,
we will focus on cat maps, which are linear automorphisms of the torus that 
exhibit hard chaos. Anosov's theorem establishes that they are structurally 
stable, that is that orbits of a slightly perturbed map are conjugated to those of the unperturbed map by a homeomorphism. To make the paper self contained we present in this section the main aspects of the classical 
and quantum perturbed cat maps that are important for our study.

\subsection{Classical system}

Cat maps for a two dimensional torus considered in a unit periodic 
square are represented by matrices acting on the coordinates:

\begin{equation}
\begin{bmatrix}
q' \\
p' 
\end{bmatrix}
=
\begin{bmatrix}

          g_{11}   & g_{12} \\
          g_{21}   & g_{22} \\
\end{bmatrix}
\begin{bmatrix}
q\\
p
\end {bmatrix}
\mod 1.
\label{cat}
\end{equation}

We take integer entries in the matrix $G$ to ensure continuity, and  also $ Tr[G]>2$ and $\det[G]=1$ since the map is hyperbolic and conservative. 

The maximal logarithm of the eigenvalues of $G$ defines its Lyapunov exponent, quantity that characterizes the rate of separation of infinitesimally close trajectories.  We consider the following
cat maps,

\begin{equation}
\quad G_1=\Bigl(\begin{matrix} 2 & 1 \\ 1 & 1 \end{matrix}\Bigr)  \;\;  \quad G_2=\Bigl(\begin{matrix} 2 & 1 \\ 3 & 2\end {matrix}\Bigr) \;\;\quad G_3=\Bigl(\begin{matrix} 4 & 1 \\ 15 & 4 \end{matrix}\Bigr).
\label{cat2}
\end{equation}

The corresponding Lyapunov exponents are $\lambda_1=\log(\frac{1}{2}(3+\sqrt5))\approx 0.96$, $\lambda_2=\log(2+\sqrt3))\approx 1.32$ and $\lambda_3=\log(4+\sqrt15) )\approx 2.06$ respectively. 

Now  we  introduce a perturbation of the cat map that is a shear in momentum that depends on coordinates and we take its linear part equal to zero at the origin, according to \cite{Ozorio 1994},  

\begin{equation}
\begin{bmatrix}
q' \\
p' 
\end{bmatrix}
=
G
\begin{bmatrix}
q\\
p+\epsilon(q)
\end {bmatrix},
\label{pert1}
\end{equation}
where in particular, we considered,
 
\begin{equation}
\epsilon (q) =\frac{k}{2\pi}(cos(2\pi q)-cos(4\pi q)),
\label{pert11}
\end{equation}
being $k$ the strength of the perturbation. We note that $k<0.11$ for the perturbation strength to satisfy the Anosov theorem \cite{Ozorio 1994}.  We have also used a more general perturbation that is a shear
in momentum and position (see below).  We notice that the Lyapunov exponent of all the cat maps 
that we have considered do not change significantly when the mentioned perturbations are taken into 
account.

\subsection{Quantum system}

Quantization of perturbed cat maps  has been an important step in the quantum chaos studies 
because these simple systems allow the understanding of  many manifestations 
of chaos in quantum systems \cite{Hannay 1980, Ozorio 1994,Espositi 2005}.   
The wave function should  be periodic in both position and momentum representation
due to  the periodic nature of the torus. Consequently, the wave functions are periodic 
combinations of delta functions. If in the coordinate representation the wave function 
has a period  $\Delta q$ with spacing $\Delta q/N$, then in the momentum representation 
the period is $\Delta p=2\pi \hbar N/\Delta q$ with spacing $2\pi\hbar/\Delta q$. If  
$\Delta p=\Delta q=1$, it follows that $1=2\pi\hbar N$. Then, we have a Hilbert space of $N$ 
dimension for a fixed value of $\hbar$. As $N$ takes increasing values, we reach the semiclassical limit.   

The propagator is obtained from the quadratic generating function and
it can be written as a $N \times N$ matrix acting over a vector with components being the amplitudes of each delta function. Up to a phase, it results,

\begin{equation}
U^{G}(q',q)=\sqrt{\frac{N}{ig_{12}}} exp \left[ \frac{i\pi N}{g_{12}}(g_{11}q^{2}-2q'q+g_{22}q'^{2}) \right].
\label{pertq0}
\end{equation}

The phase factor is equal to the unity if $g_{12}=1$. It can be shown that only cat maps with odd diagonal and even antidiagonal (or visceversa) can be quantized \cite{Hannay 1980}. In the case of $G_1$ the number $N$ of states of the Hilbert space must be even \cite{ford91}, requirement that  we have satisfied.

In order to quantize the perturbed map, we use the fact that it can be written as a composite map.  
If P is the simple shear, the quantized map is thus defined by composing the evolution operators for 
the cat map and the shear in the same order as the classical propagator:

\begin{equation}
U=U^{G}U^{P},
\end{equation}
being $U^P=exp[i2\pi N S_p(q)]$ and $\frac{-dS_p(q)}{dq}=\epsilon(q)$, so,

\begin{equation}
S_p(q)=\frac{k}{4\pi^{2}}(sin(2\pi q)-\frac{1}{2}sin(4\pi q)).
\label{accion}
\end{equation}

Now we define the scaled strength,

\begin{equation}
\chi \equiv \frac{k}{2 \pi \hbar} =k N. 
\end{equation}

Finally, we have the propagator for the perturbed cat map,

{\setlength\arraycolsep{1pt}
\begin{eqnarray}
\ U(q',q)=\sqrt{\frac{N}{ig_{12}}} exp \left[ \frac{i\pi N}{g_{12}}(g_{11}q^{2}-2q'q+g_{22}q'^{2}) \right]
{}\nonumber\\
{}exp \left[ \frac{i \chi }{2\pi}(sin(2\pi q)-\frac{1}{2}sin(4\pi q)) \right].
\label{pertq1}
\end{eqnarray}}

To introduce more general perturbations, we consider shears in both momentum and positions \cite{Espositi 2005}. Eq. (\ref{pertq1}) shows that a shear in momentum is diagonal in the coordinates representation. 
For a shear in position, we can construct the propagator by changing to momentum coordinates and finding a diagonal matrix in this representation. The change of basis matrix is the discrete 
Fourier transform $F$, where,

\begin{equation}
F_{lj}=\frac{1}{\sqrt{N}} exp \left( \frac{-2\pi ilj}{N} \right).
\end{equation}

In this work, we use the following shears in momentum and positions:

\begin{equation}
U=U^{P}U^{G}F^{+}U^{Q}F,
\label{pertq2}
\end{equation}
where $U^{Q}=exp[\frac{i\chi}{2\pi}(cos(6\pi p)-\frac{1}{2}sin(4\pi p))]$.

\subsection{Local Density of states}
The action of a perturbation on the eigenstate  of a quantum system can be described by the LDOS.
Let $\phi_j(k)$ and $\psi _j(k)$ be the eigenphases and eigenfunctions of a perturbed cat map ($k$
is the strength of the perturbation and $j=1,...,N$).  For an eigenstate $i$ at $k_0$, considered as the unperturbed eigenstate, the LDOS 
is defined as follows,

\begin{equation}
\rho_i(\phi,\Delta k)= \sum_{j} \vert\langle \psi _j (k)\vert \psi _i (k_o)\rangle\vert^{2} \delta (\phi -[\phi_j(k)-\phi_i(k_0)])
\label{LDOS}
\end{equation}
where $\Delta k=k-k_o$. The  initial strength $k_0 \neq 0$ is needed to remove the peculiar non generic behavior of the unperturbed cat map  \cite{Hannay 1980, Ozorio 1994}. To avoid a dependence on the particular characteristics of the unperturbed state $i$ we always make an average over this state. 

The width $\sigma$ of the LDOS gives us an idea of how many states in the base of the unperturbed system 
are needed to describe a state of the perturbed one. So, it is a good measure of the perturbation action 
over the system. This quantity can be measured in different ways \cite{wisniacki3}. 
In our case, we take the half distance around the mean value that contains the $70 \%$ of the probability \cite{wisniacki3}.   

\begin{figure}[h]
\setlength{\unitlength}{1cm}
\begin{center}
\includegraphics[width=8.5cm]{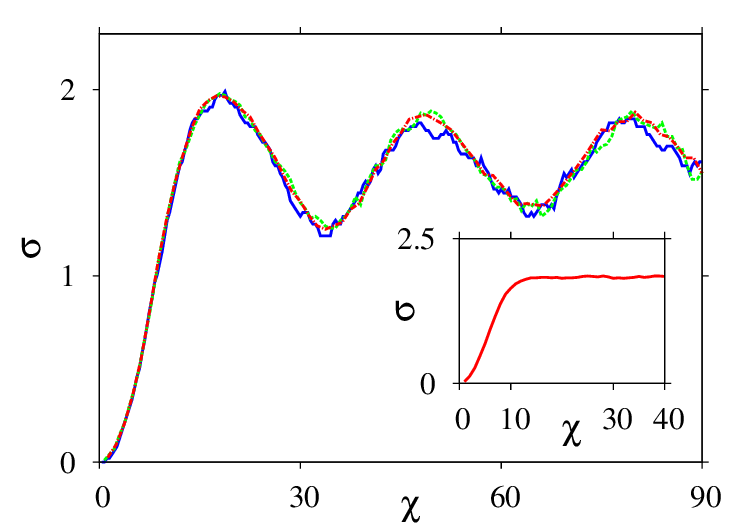}
\end{center}
\caption{(Color online) Width $\sigma$ of LDOS as a function of the scaled perturbation strength
$\chi$ for $N=300$ (green line), $N=600$ (blue line) and $N=1200$ (red line). The perturbation is
a shear in positions for the cat map given by $G_2$. We have obtained the same results for $G_1$ and $G_3$. 
It is clear that $\sigma$ does not depend on the size of the Hilbert space and exhibits an oscillatory 
behavior for strong perturbations.Inset: $\sigma$ vs. $\chi$ for the same cat map perturbed with shears 
in both 
momentum and positions.  It remains steady for strong perturbations.}
\label{fig1}
\end{figure}

\begin{figure}[h]
\setlength{\unitlength}{1cm}
\begin{center}
\includegraphics[width=8.5cm]{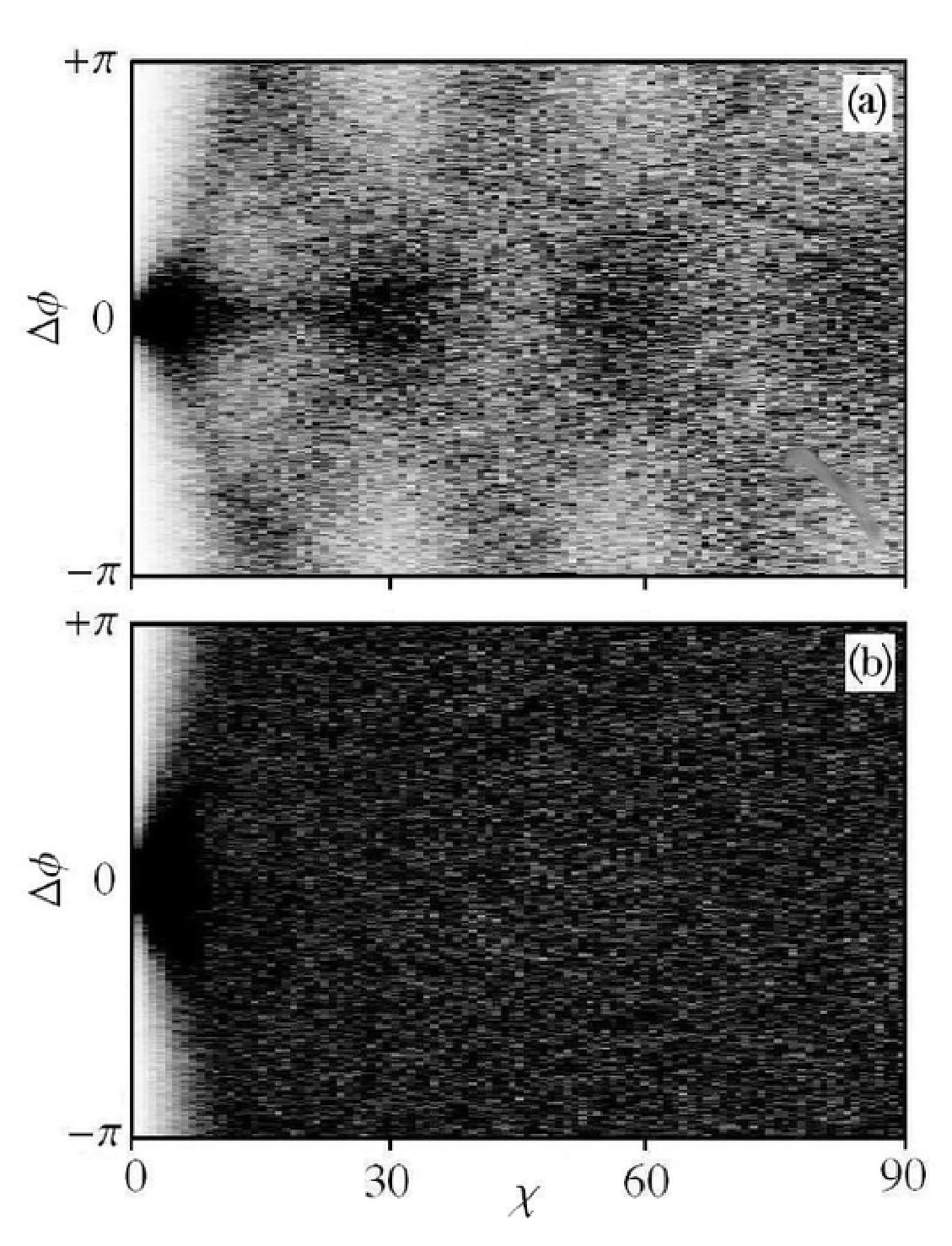}
\end{center}
\caption{Density plots of the LDOS $\rho(\Delta \phi, \chi)$  as a function of $\chi$. An average over 50 
unperturbed states was done.  (a) The perturbation is a simple shear in momentum. We can notice the coherence that is responsible for the oscillations found in the width of the LDOS. (b) The shear is in momentum and coordinates. In this case, no coherence is observed.}
\label{fig2}
\end{figure}

We compute the width $\sigma$ for both perturbations introduced before (see Fig.\ref{fig1}). 
In the main plot of Fig.\ref{fig1}, $\sigma$ is shown as a function of $\chi$ when the cat map
$G_2$ is perturbed  with a  shear in momentum [Eq. \ref{pertq1}].  The initial value of the  strength is  $k_0=0.02$ and the width is computed for $N=300$, $N=600$ and $N=1200$. We clearly see
that the width of the LDOS does not depend on the number of states of the Hilbert space.
We also notice that $\sigma$  has an oscillatory behavior for $\chi \gtrsim 15 $. In the small perturbation
regime ($\chi \lesssim 15 $), we see the quadratic behavior 
that can be obtained in a perturbative way \cite{doron1}. 
For shears in momentum and position [Eq. (\ref{pertq2})], we also compute  $\sigma$ as a function of 
$\chi$ for $N=1200$ [see the inset of Fig. (\ref{fig1})]. In contrast to the previous results, the width 
$\sigma$ is constant for $\chi \gtrsim 15 $. The quadratic perturbative regime is present again for small perturbations. 

From the results of Fig. (\ref{fig1}), it is clear that the oscillatory behavior is related to the type of perturbation considered. In order to illustrate this behavior in more detail, we show the LDOS 
[Eq. \ref{LDOS}] as a function of the scaled strength $\chi$ for the considered perturbations. 
We have made an average over 50 unperturbed states.
Fig.\ref{fig2} (a) is a density plot of  $\rho(\Delta \phi, \chi)$  when the cat map $G_2$ is perturbed with a simple shear in momentum  [Eq. (\ref{pertq1})] and, in Fig. \ref{fig2} (b), with both shears in momentum and positions [Eq. (\ref{pertq2})]. In the first case [see Fig. \ref{fig2} (a)] and for $\chi < 10 $, the distribution has a big peak with Lorentzian shape whose width becomes wider until its tails reach the limits of the Hilbert space . Then two small peaks appear at $\chi \approx 10$. These peaks move to the border 
and then come back to the center from both sides. That denotes the coherence given by this kind of perturbation. In contrast, when the shear is more general, the distribution is almost uniform for 
$\chi > 10$ and it lacks any coherence. We remark that the results that are shown in Figs. \ref{fig1} and \ref{fig2} for
$G_2$ are almost the same for  the maps $G_1$ and $G_3$.

\begin{figure}[h]
\setlength{\unitlength}{1cm}
\begin{center}
\includegraphics[width=9.5cm]{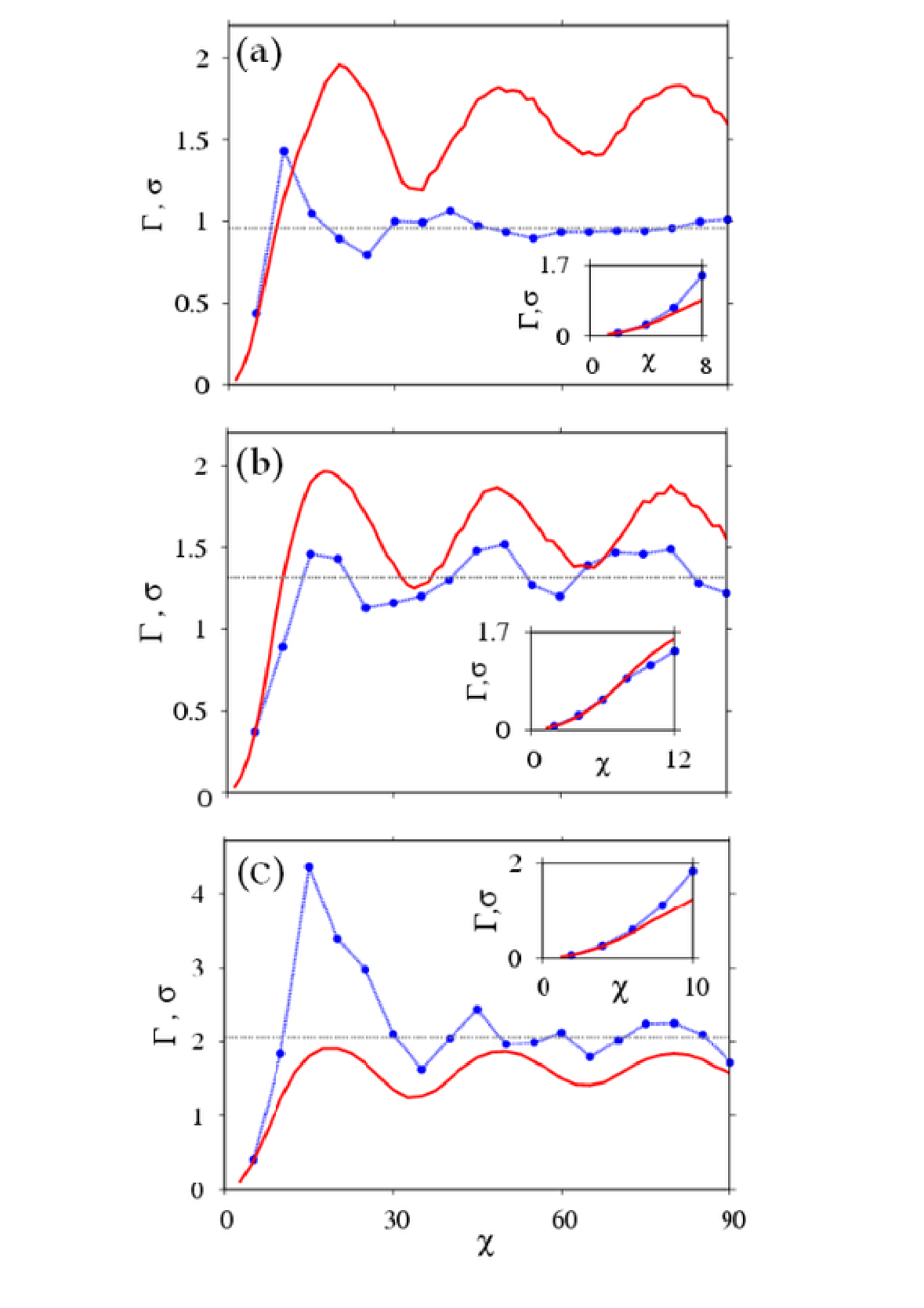}
\end{center}
\caption{(Color online) Decay rate  $\Gamma$ of the LE (full circles)  of a cat map perturbed with a shear in momentum as a function of the scaled perturbation strength $\chi$.  The width $\sigma$ of the LDOS
 (solid lines) and the  Lyapunov exponent (dotted lines) are also plotted.  
(a) The cat map is given by $G_1$. In this case  $\lambda_1 < \sigma$ for $\chi>15$. Here the LDOS  do not influence the LE decay in the strong perturbation regime. (b)  The cat map is given by $G_2$. In this case $\lambda_2$ is similar to the strong perturbation limit of $\sigma$. Now we clearly see that the oscillations of the LE decay rate are strongly related to the ones of $\sigma$. (c)  The cat map is given by $G_3$. In this case  $\lambda_3 > \sigma$  for strong perturbations. The oscillations exhibited by the LE decay rate follow the $\sigma$ ones with a larger amplitude. In the insets the weak perturbation regime can be clearly seen (FGR regime).}
\label{fig3}
\end{figure}
\section{Loschmidt Echo}
\label{sec:LE} 

This section is devoted to the study of the decay rate $\Gamma$ of the exponential decay of the LE.
We compute $\Gamma$ as a function of $\chi$ for the cat maps $G_1$, $G_2$ 
and $G_3$ perturbed by a shear in momentum [Eq. \ref{pertq1}]  and by shears in momentum and position
[Eq. \ref{pertq2}]. 
We have used $N=2000$ in all the computations of this Section. This value of the number of states of the Hilbert space allows for a 
good fit of the decay of the LE due to the fact that the long-time saturation occurs at a value inversely proportional to the  effective size of the Hilbert space of the system.  The initial states are 
coherent states randomly localized in phase space. An average over 200 initial states was made.

\begin{figure}[h]
\setlength{\unitlength}{1cm}
\begin{center}
\includegraphics[width=8.5cm]{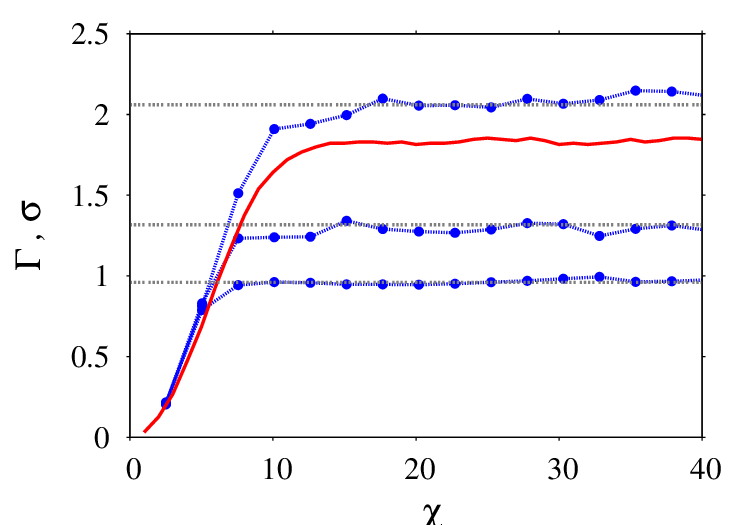}
\end{center}
\caption{(Color online) Decay rate  $\Gamma$ of the LE (full circles)  of a cat map perturbed by a shear in momentum and position as a function of the scaled perturbation strength $\chi$.  The width $\sigma$ of the LDOS (solid lines) and the  Lyapunov exponent (dotted lines) are also plotted.   The cat map is given by $G_1$ (bottom),  $G_2$ (middle) and $G_3$ (top). See text for details.}
\label{fig4}
\end{figure}

Fig. \ref{fig3} shows $\Gamma$ as a function of $\chi$ for $G_1$ [panel (a)], $G_2$ [panel (b)], and $G_3$ [panel (c)] perturbed by a shear in momentum.  The width $\sigma$ of the LDOS (solid lines) and the value of the Lyapunov exponents (dotted lines) are also plotted.  In the insets we show the behavior
of $\Gamma$ in the weak perturbation regime. It is clear that for weak perturbations the decay rate of 
the LE and the width of the local density of states are equal: both of them follow the quadratic 
behavior given by the Fermi Golden Rule. For strong perturbations, $\Gamma$ reaches the value of the Lyapunov exponent in the case of $G_1$, when the  oscillations of the width of LDOS occur at greater value than $\lambda_1$[Fig.\ref{fig3} (a)].
However, in the last two cases, it exhibits an oscillatory behavior that is evidently related to $\sigma$ [Fig.\ref{fig3}(b) and (c)]. In the case of the map given by $G_2$, $\lambda_2$ is less than the limit reached by $\sigma$ and we have just mentioned that we can observe the influence of the width of the local density of states in $\Gamma$. For $G_3$, $\lambda_3$ is greater than that limit, and the influence seems to be stronger. In this case, we see in Fig. \ref{fig3} (c) that the decay $\Gamma$ near the first peak of $\sigma$ at $\chi \approx 15$ is strongly enhanced and the decay rate reach to  $\Gamma \approx 2 \lambda$.   The $2 \lambda$ decay was studied using the uniform semiclassical approach in Ref. \cite{wang}. This behavior deserves further investigation \cite{Ares2}. 

However, for the cat map given by $G_3$, there are some points for $45<\chi<60$ that seem to be misplaced. The reason for this disagreement is that the fast decrease of the echo caused by the large Lyapunov exponent makes $\Gamma$ very difficult to determine. As we want to verify the behavior found for $\Gamma$ when $\lambda$ is greater than the limit reached by $\sigma$, we repeated the calculations for the cat map given by,

\begin{equation}
\quad G_4=\Bigl(\begin{matrix} 8 & 1 \\ 63 & 8 \end{matrix}\Bigr),
\label{cat3}
\end{equation}
whose Lyapunov exponent is $\lambda_4=\log(8+3\sqrt7) )\approx 2.77$. The results were completely analogous. To improve the fitting  it would be necessary to run these calculations for greater values of $N$, so the decay rate could be better distinguished but it would demand a really long computational time
because the computational time goes as $N^2$. In the next section we will show that local perturbations allow to study the case in which the Lyapunov exponent is much smaller that the width of the LDOS.


In order to confirm that the oscillatory behavior observed in the decay rate of the LE is due  to the influence the oscillations of the width of LDOS, we also make the same study as before for the perturbation [Eq. (\ref{pertq2})] that does not have any oscillations [see the inset of Fig. \ref{fig1} and Fig. \ref{fig2}(b)].
In Fig. \ref{fig5} we show $\Gamma$ for $G_1$, $G_2$ and $G_3$, together with $\sigma$, for this general perturbation.
It is clear that as $\sigma$ has no oscillations, nor does $\Gamma$, and this behavior is the same for the different cat maps considered. In all cases, they reach the Lyapunov value for strong perturbations. The last observation supports our conclusion about the influence of the  oscillations of the width of LDOS over the decay of the LE in the Lyapunov regime.

Now, we can go one step further by giving a semiclassical interpretation of our results using the dephasing representation of the LE by Vanicek \cite{Vanicek 2003}. The LE is found to be,
  
{\setlength\arraycolsep{1pt}
\begin{eqnarray}
M(t)=\vert O(t)\vert^{2}=\vert (\frac{A^{2}}{\pi\hbar^2})^{N/2} \int d^{N}p' exp[\frac{i}{\hbar}
{}\nonumber\\
{}\Delta S_s(r,r_0,t)-(p'-p_0)² \frac{A²}{\hbar²}]\vert^{2}
\end{eqnarray}}
\noindent where $r_0$ is the center of the initial Gaussian wave packet, with dispersion $A$ and an average momentum $p_0$. $\Delta S_s(r,r_0,t)$ stands for the difference of action of the perturbed and unperturbed orbits.
Now, when the shear caused by the perturbation is an integer number of $1/N$ (the grid of the quantum phase space) the interference between the perturbed and the unperturbed orbits, given by the difference of action, is constructive. Otherwise, it is destructive. So, when the shear is in momentum, that condition is satisfied for certain values of $k$ and consequently we observed oscillations with a wave length proportional to $1/N$. 
The wave length of the oscillations of $\sigma$ would be, therefore, $4\pi^2/N$ in units of $k$ [Eq. (\ref{accion})], as we can verify in Fig.\ref{fig1}, where it is approximately $4 \pi^2$ in units of $\chi$. 

On the contrary, when the shear is in momentum and positions, the interference condition can not be satisfied simultaneously and the oscillations disappear.

\section{Loschmidt echo for local perturbations}
\label{sec:LELP}

Perturbations to quantum systems are generally local when they come from experiments. 
This means that understanding of the general behavior of LE 
for local perturbation is an important issue. Recently,  Goussev and co-workers have presented a 
semiclassical theory for the decay of the LE for local perturbations in billiards systems 
\cite{Richter 2008}.   
They show that this kind of perturbations introduce an oscillatory behavior of the  decay of the LE  around the classical escape rate. This quantity corresponds to the probability to escape from the 
billiard when the classical particle hits the region of the  boundary  where the perturbation is applied. 

\begin{figure}[h]
\setlength{\unitlength}{1cm}
\begin{center}
\includegraphics[width=8.7cm]{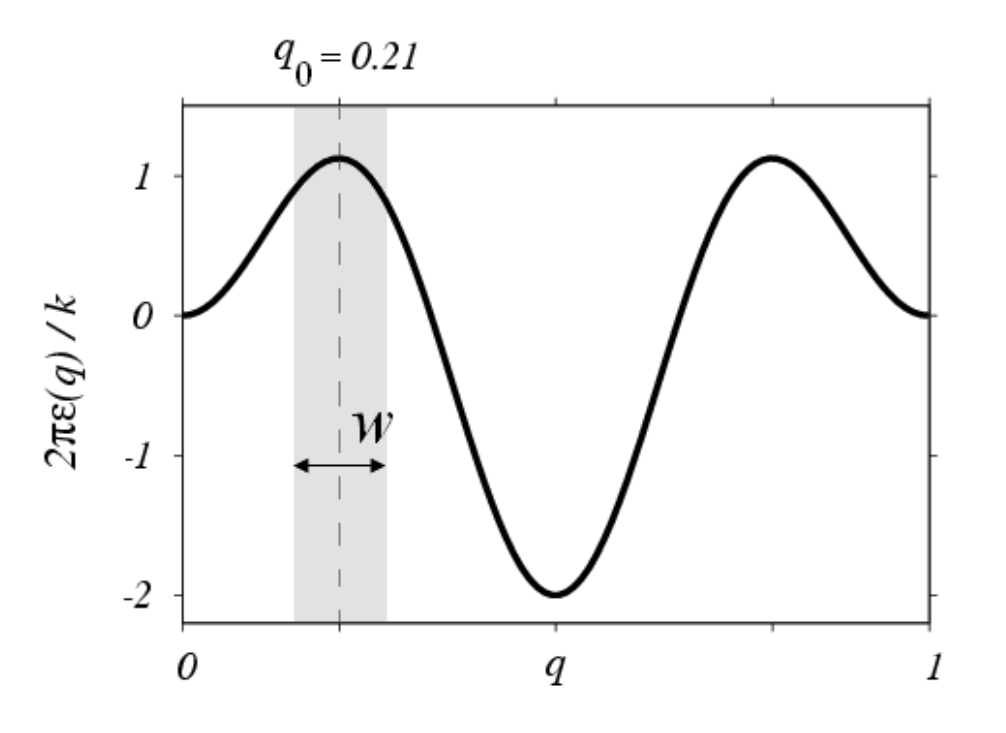}
\end{center}
\caption{Schematic figure showing  the local perturbation that was used. The scaled 
shear [$2 \pi \epsilon(q)/k$] ($k$ is the strength of the perturbation) is plotted as a function of the $q$ 
coordinate. $q_0$ indicates the center of the interval that was perturbed and $w$ its width.}
\label{fig5}
\end{figure}

So far, we have shown in previous sections that the characteristics of the perturbation modify the
behavior of the  width of the LDOS and that they shape the decay of the LE.  With this in mind we have looked for that dependence in the case of local perturbations. Furthermore, we are interested in the behavior of the decay rate of the LE
as a function of the width of the perturbed region.
With this aim, we have applied the already introduced shear in momentum to a window of coordinates of the phase space. We took the coordinate in which the function that describes the shear $\epsilon(q)$ [Eq. \ref{pert11}] has one of its maxima  as the center of the interval to be perturbed. This is shown in Fig.\ref{fig5}, in which
we plot the scaled $\epsilon(q)$ and the corresponding interval that is perturbed.  

\begin{figure}[t]
\setlength{\unitlength}{1cm}
\begin{center}
\includegraphics[width=8.5cm]{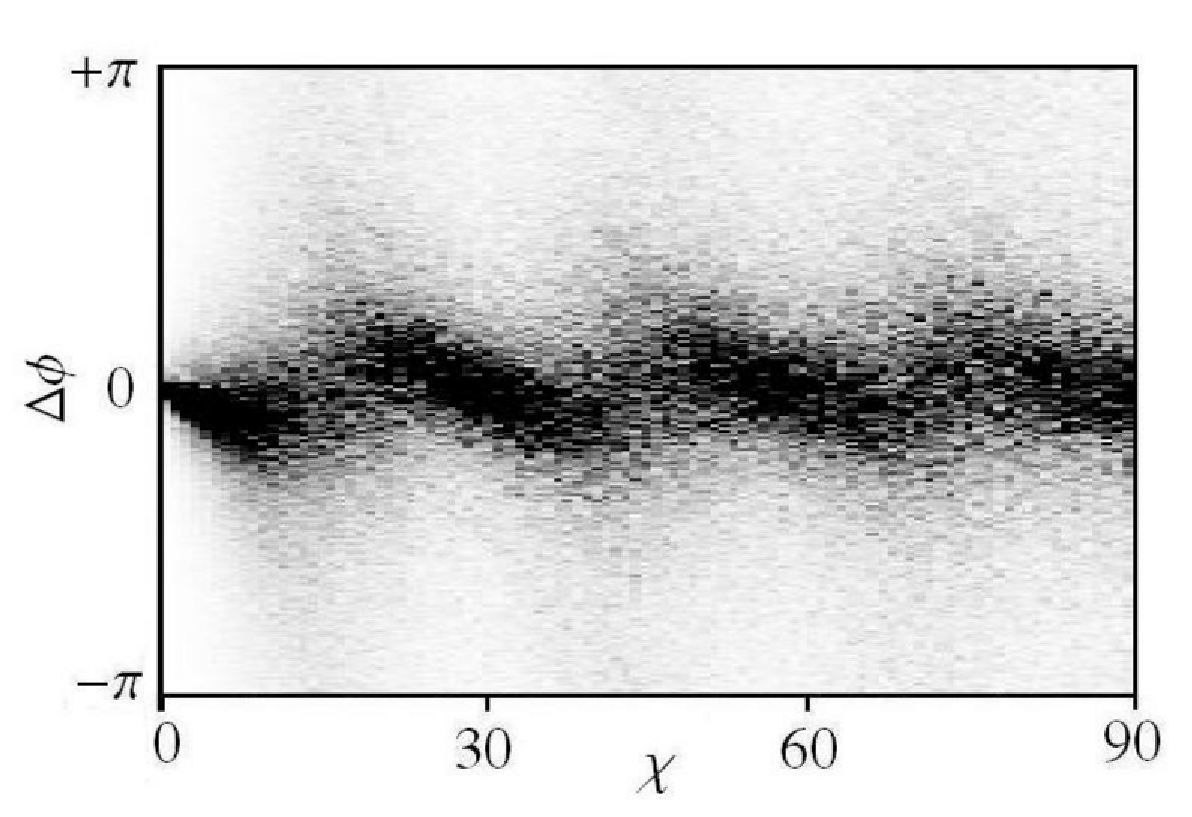}
\end{center}
\caption{Density plot of the LDOS $\rho(\Delta \phi, \chi)$  as a function of the strength $\chi$ when the perturbation is a simple shear in momentum applied in a region of width $w=0.4$ (see Fig. 5).
It can be clearly  seen that it is restricted to the perturbed phases and that it has a coherent behavior.}
\label{fig6}
\end{figure}

Let us first see the influence of a local perturbation in the behavior of LDOS. In Fig.\ref{fig6} we exhibit the distribution $\rho(\Delta \phi, \chi)$ when the perturbation is applied in a region of $w=0.4$. It is clear that the distribution is bounded and we can see a  
coherent behavior similar to the one observed in Fig.\ref{fig2} (a).
We notice that  $\rho(\Delta \phi, \chi)$ does not present the same symmetry as around $\Delta \phi=0$ as is
observed in  Fig.\ref{fig2} (a) when the perturbation is global. This is related with the position $q_0$ as 
center of the perturbed region \cite{Ares2}.

\begin{figure}[t]
\setlength{\unitlength}{1cm}
\begin{center}
\includegraphics[width=8.5cm]{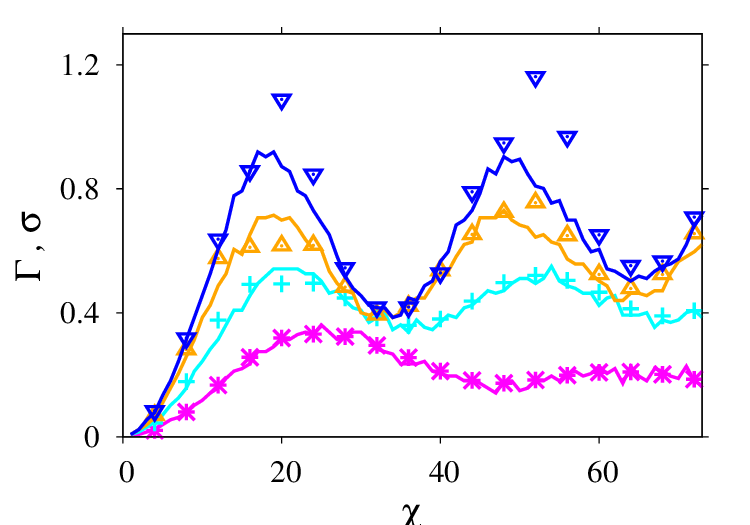}
\end{center}
\caption{(Color online)
Decay  $\Gamma$ of the LE of a cat map perturbed locally with a shear in momentum as a function of the scaled perturbation strength $\chi$. The  symbols correspond to the following width of the perturbed region:  $\bigtriangledown$  ($w=0.4$),  $\bigtriangleup$  ($w=0.3$),  $+$ ($w=0.2$) and $\ast$ ($w=0.1$). The width $\sigma$ of the LDOS   is also plotted with solid lines.  
It is noticeable that, the smaller is $w$, the more similar to $\sigma$ $\Gamma$ is. In all the cases considered, the oscillations are equal for both magnitudes.}
\label{fig7}
\end{figure}

\begin{figure}[t]
\setlength{\unitlength}{1cm}
\begin{center}
\includegraphics[width=8.5cm]{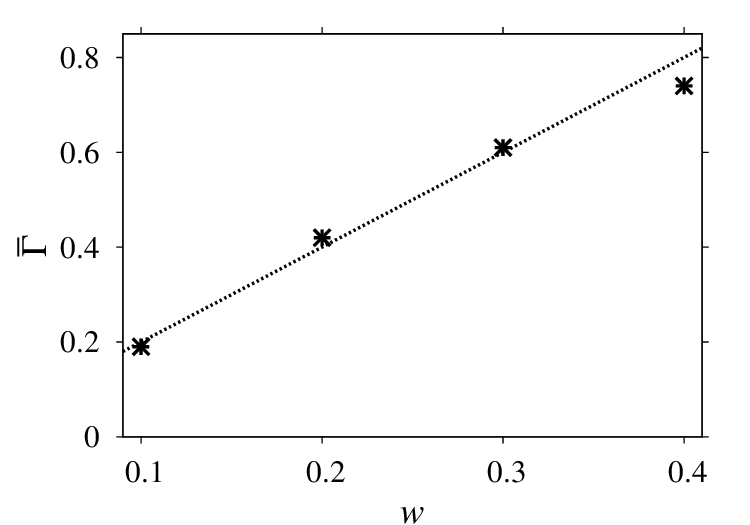}
\end{center}
\caption{Mean value $\bar{\Gamma}$ of decay rate of the LE as a function of the width $w$ of the applied 
local perturbation. The function $2 w$ is also plotted with dotted line . See text for details.}
\label{fig8}
\end{figure}

In Figs. \ref{fig7} and Fig.\ref{fig8} we show the behavior of the decay of the LE for a local perturbation. All  the computations in this section were done with $N=800$ (the total size of the Hilbert space). 
We have
considered several widths where the perturbation was applied. As an example, in Fig.\ref{fig7} we show the decay $\Gamma$ of the LE  with symbols for:  $\bigtriangledown$  ($w=0.4$),  $\bigtriangleup$  ($w=0.3$),  $+$ ($w=0.2$) and $\ast$ ($w=0.1$). We notice that, although the oscillatory behavior is present for all the values of $w$ considered, its amplitude is smaller when $w$ decreases. Moreover, the value around the oscillation takes place also becomes smaller as $w$ becomes thinner.
To conclude the analysis, we also compute the width $\sigma$ of the LDOS for all the different apertures considered. The width $\sigma$  is shown in Fig.\ref{fig7}  with solid lines. 
It is clear that for $w=0.1$ and $w=0.2$, the behavior of $\Gamma$ and $\sigma$ are identical. 
That is, the decay of the LE is given by the width of the LDOS.  This result is in agreement with the semiclassical study for billiard systems  of Ref.  \cite{Richter 2008}.
For bigger $w$ ($w=0.3$ and $w=0.4$ in our case), the oscillations are alike, but its amplitude is not so similar.   So, in the case of a local perturbation, the influence of the LDOS width over $\Gamma$ is enhanced. 

In Fig.\ref{fig8} we show the mean value of the decay of the LE $\bar{\Gamma} $ after the quadratic growth [$\chi>20$, see  Fig. \ref{fig7}] as a function of the width $w$ of the perturbation. We can clearly see that  $\bar{ \Gamma}  \approx 2 w $ for $w<0.4$. We remark that our result agrees
with the one obtained for billiard systems in Ref. \cite{Richter 2008} because the width 
$w$ corresponds to the classical escape rate in our system. \\

\section{Final remarks}
\label{sec:FR} 

In this work we have studied the decay of the LE for strong perturbations, that is,  
the regime after the quadratic growth of the decay  (FGR regime). 
We have shown that the expected monotonic regime in which the decay is given by the
Lyapunov exponent of the classical system  is not
universal. In fact, the behavior of the LE in this regime depends on the ratio between the
mean value of the width of LDOS $\bar{\sigma}$ and the  value of the Lyapunov exponent.

The general idea is the following. When a system is perturbed and the perturbation affects a finite 
number of states of the Hilbert space, instead of growing monotonically, the width of the LDOS 
reaches a plateau $\bar{\sigma}$ or an non-uniform regime around $\bar{\sigma}$. If $\bar{\sigma}/  \lambda>>1$,   the decay rate of the LE is given by  $\lambda$ and does not depend on the perturbation.  
But, if  $\bar{\sigma}/\lambda \leq 1$ and  the width $\sigma$ of the LDOS has
a non uniform behavior,  a non-monotonicity is imprinted in the decay of 
the LE. This regime was recently  reported in the literature 
\cite{coldatoms,Casati 2004,Li 2005, Pozzo 2007} but not the conditions for it to happen.  

We have illustrated the previous ideas using  cat maps with different Lyapunov  exponents and perturbations. When the cat map is perturbed with a simple shear in momentum, we have found pronounced oscillations of the width of the LDOS, that are a consequence of the coherence of perturbed and unperturbed orbits that this kind of perturbation introduces.
Significantly, we have shown that this oscillating regime was also present in the decay rate of the LE when the value around the LDOS width oscillates is close enough to  the Lyapunov exponent of the classical system. Moreover, when the perturbation is more general, the coherence between orbits is destroyed, 
so the width of the  LDOS reaches a plateau and so does the LE decay.

We have also studied the decay  of the LE in cat maps when the perturbation is local \cite{Richter 2008}. 
The important point in this case is that varying the width of the region in which the perturbation is 
applied we can change the width of the LDOS while the Lyapunov exponent remains fixed.
 With this in mind, we have perturbed a range of coordinates with a simple shear in momentum. We have found that the decay
rate  is almost identical to the width of the LDOS when the range of  the perturbed coordinates is 
very small. Moreover, we have shown that the amplitude of the oscillations of the LDOS decreases
with the width of the applied perturbation and the mean value is given by the classical decay rate.


\section{Acknowledgements}

The authors acknowledge the support from CONICET (PIP-6137) , UBACyT (X237) and ANPCyT.  D.A.W. are researcher of CONICET.  We would like to thanks Ignacio Garc\'{\i}a Mata, Pablo Tamborenea and Eduardo Vergini for
useful discussions.



\end{document}